\documentclass[]{interact}

\usepackage{epstopdf}

\usepackage[font=footnotesize]{caption}
\DeclareCaptionLabelSeparator{None}{ }
\usepackage[font=footnotesize]{subcaption}
\usepackage{multicol}
\usepackage{hyperref}
\usepackage{wrapfig}
\usepackage{fancyhdr}
\fancypagestyle{title}{
	\fancyhf{}
	\lhead{\textit{Combustion Theory and Modelling, 2019}}
	\rhead{}
}

\usepackage[numbers,sort&compress]{natbib}
\bibpunct[, ]{[}{]}{,}{n}{,}{,}
\renewcommand\bibfont{\fontsize{10}{12}\selectfont}

\theoremstyle{plain}

\theoremstyle{definition}

\theoremstyle{remark}

\begin{document}


\title{\vspace{-2.0cm}Multiplicity of detonation regimes in systems with a multi-peaked thermicity}

\author{
\name{S. SM. Lau-Chapdelaine\textsuperscript{a}\thanks{Corresponding
author: S.~SM. Lau-Chapdelaine. Email: slauc076@uottawa.ca}, F. Zhang\textsuperscript{b}, and M.I. Radulescu\textsuperscript{a}}
\affil{\textsuperscript{a}Department of Mechanical Engineering, University of Ottawa, Ottawa, Ontario, Canada, K1N 6N5; \textsuperscript{b}Defence Research and Development Canada, Medicine Hat, Alberta, Canada, T1A 8K6}
}

\maketitle\thispagestyle{title}

\begin{abstract}
The study investigates detonations with multiple quasi-steady velocities that have been observed in the past in systems with multi-peaked thermicity, using Fickett's detonation analogue. A steady state analysis of the travelling wave predicts multiple states, however, all but the one with the highest velocity develop a singularity after the sonic point. Simulations show singularities are associated with a shock wave which overtakes all sonic points, establishing a detonation travelling at the highest of the predicted velocities.

Under a certain parameter range, the steady-state detonation can have multiple sonic points and solutions. Embedded shocks can exist behind sonic points, where they link the weak and strong solutions. Sonic points whose characteristics do not diverge are found to be unstable, and to be the source of the embedded shocks. Numerical simulations show that these shocks are only quasi-stable. This is believed to be due in part to a feature of the model which permits shocks anywhere behind a sonic point.

\end{abstract}


\section{\label{sec:Introduction}Introduction}

The decomposition of certain reactive materials can occur in two or more distinct steps, characterized by multiple peaks in the thermicity (effective rate of energy release). Nitromethane-air detonations \cite{presles1996gaseous} and other usual fuels using NO$_x$ as oxidizer \cite{joubert2008detonation} give rise to such multiple reaction zone detonation structures. Thermo-nuclear fusion reactions also occur in sequential steps. Detonations in degenerate white dwarfs undergoing supernova explosions of the Type Ia have three sequential steps where carbon, oxygen, and silicon undergo fusion \cite{gamezo1999multilevel}. Hybrid detonations are self-sustained detonation waves in reactive gas mixtures with suspended reactive dust and display two sequential reaction zones in the detonation structure \cite{zhang2006detonation,afanasieva1983multifront,khasainov1988steady}. The gas phase reaction first proceeds with influence from the solid phase, including energy used to heat the particles, and momentum lost to the solid phase by entraining the particles with the gas flow. The solid phase reacts exothermically once it has absorbed sufficient energy.

A common feature of multi-peaked thermicity \cite{fickettdavis1979detonation} systems is the presence of endothermic processes coupling the multiple reactions. These losses can be manifested by heat and momentum losses to confining walls, mass divergence, or curved geometries. Losses can also be intrinsic to the system, as they are in hybrid detonations, for example, where particle heating and drag withdraw energy from the gas phase. Experiments and numerical simulations in these hybrid systems have shown that the selection rules and detonation wave pressure profiles depend intimately on the kinetics of the reactions in either phase, the amount of energy release, and hydrodynamic resistance of the particles to the gas phase motion; Zhang's recent review \cite{zhang2009detonation} provides the state-of-the-art.

In systems with simultaneous exothermic and endothermic processes, the competition between energy addition and loss dictate the structure of the self-sustained wave \cite{zeldovich1960theory}. Self-sustained, steady travelling detonations with losses have a surface of zero net thermicity within their reaction zone, where energy release is balanced by losses. This surface is sonic with respect to the detonation front, in order to avoid a singularity in the solution, and information beyond it is unable to reach the front. The detonation's velocity is therefore only influenced by the accumulated energy release between the lead shock and this sonic surface. The conditions governing the propagation velocity of detonations with losses are known as the ``generalized Chapman-Jouguet (CJ)'' conditions and are treated at length in most detonation textbooks \cite{fickettdavis1979detonation,lee2008detonation}.

In systems with multiple thermicity peaks, it is plausible that the generalized CJ condition be met multiple times, and it is unclear what governs the speed of the lead shock. Previous investigations have examined the solutions of detonations with multiple exothermic reactions and simultaneous losses.

Veyssi\`{e}re and Khasainov \cite{veyssiere1991model,veyssiere1995structure,khasainov1996initiation} numerically studied steady hybrid detonations using two-phase reactive Euler equations. They found three steady propagation regimes. One regime was driven solely by the gas reaction while particles remained inert in the driving region. The second regime had both gas and particle reactions driving the detonation front, consequently travelling faster than previous regime. Finally, the third regime was steady only under certain parameters. At these specific parameters, the detonation propagated at the same velocity as the first regime, however, they found a shock wave embedded within the reaction zone. For other sets of parameters, the embedded shock eventually overtook the detonation front and the velocity increased to that of the second regime, where reactions from the gas and solid phases drove the detonation together.

Bdzil {et al.} \cite{bdzil2006higher} investigated a two-reaction case with losses due to shock curvature, using asymptotic expansions of the reactive Euler equations. They found a range of curvatures (losses) where two quasi-steady detonation velocities were possible. Time-evolution of a blast-initiated detonation showed that, with certain initial conditions, the blast would decay such that the flow would adopt the slower detonation velocity, then in some cases abruptly transition to the fast solution.

A multiplicity of steady velocity solutions for detonations with multi-peaked thermicity were found in these studies, some unstable. Internal shocks were found to be transient in some cases and steady in others. The selection rules for the detonation structure and velocity, however, remain unclear. This is partially because the origin and transient of the internal shocks is uncertain.

This investigation aims to clarify the steady reaction zone structure and velocity of detonations with multi-peaked thermicity. Solutions and selection rules for situations where two sonic planes propagate at different speeds will be explored. The case where the two sonic planes propagate at the same velocity is complex and poorly understood and will also be studied.  Experiments and simulations in hybrid systems \cite{zhang2006detonation,veyssiere1995structure, zhang2009detonation,afanasieva1983multifront,veyssiere1984double} have suggested that a double reaction zone structure is possible, with an embedded shock located between two sonic planes. It is presently unclear if such a shock is stable, if it depends on external losses, as modelled by previous authors, or is dependent on other three-dimensional effects in the experiments. The wave structure of such double-shock detonations, their stability, and origin of embedded shocks will be studied analytically and numerically.

The flow-fields of detonations with losses and multiple heat releases are very complex and dependant on reaction mechanisms, viscous, turbulent, and multi-dimensional effects that can be particular to specific cases. The granularity of these effects is lost in the continuum and volume-averaged limits, however, models in these limits are still viable and provide qualitative insights on the dynamics of these hyperbolic systems.
For example, Brailovsky and Sivashinsky \cite{brailovsky2000hydraulic,brailovsky2002effects}, Dionne et al. \cite{dionne2000transient}, and Semenko et al. \cite{semenko2016set} studied the multiplicity of regimes in single-reaction detonations subject to friction and convection losses. Their models were based on the one-dimensional Euler equations and assumed simple, volumetrically-averaged heat addition and losses. These continuum simplifications allowed them to qualitatively explain \cite{brailovsky2000hydraulic,brailovsky2002effects} the complex multi-valued and set-valued \cite{semenko2016set} behaviours of `low velocity' detonations in very rough or porous channels. 

In the present study, we address the structure and stability of detonations with multiple peaks of thermicity using a simplified model first introduced by Fickett \cite{fickett1979detonation} and Majda \cite{majda_qualitative_1981} as a qualitative model for reactive gas dynamics problems.  It takes the reactive form of the inviscid Burgers equation
\begin{equation*}
\frac{\partial u}{\partial t} + \frac{\partial}{\partial x} \left( \frac{1}{2} u^2 + \lambda Q \right) = 0.
\end{equation*}

It can be shown that the present model can be obtained from the reactive Euler equations in the Newtonian limit (the isentropic exponent approaching unity) and low heat release \cite{clavin_dynamics_2002,faria_theory_2015}.  It applies to transonic flows in the vicinity of sonic points of detonations \cite{bdzil1994weakly}. This makes the model directly relevant for analyzing the interplay between multiple transonic layers present in systems with multiple thermicity peaks.  Note that this limit also coincides with the conditions for weakly reactive dusty gas with high dust loading \cite{marble_dynamics_1970}. This model problem is studied due to the simplicity it offers over the full reactive Euler equations in interpreting the structure and stability of detonations with multiple peaks of thermicity. The simplicity comes from the presence of a single family of non-linear propagating waves and its coupling to the thermicity. 

In the past, this simplified model has found great utility in understanding dynamic phenomena involving detonations, such as their one-dimensional and two-dimensional stability \cite{clavin_dynamics_2002,radulescu2011nonlinear,kasimov_model_2013,kabanov2018linear}, dynamics of heterogeneous detonations \cite{mi_influence_2015}, detonation initiation \cite{lau-chapdelaine_planar_2017} and the eigenvalue structure and limits in the presence of losses \cite{fickett1985introduction,faria2015qualitative}. The present study extends this substantive body of work to reactive systems with multiple thermicity peaks. While the authors recognize the limitations of this simplified model, it is stressed that the present study serves as a stepping stone to the investigation of the full set of reactive multi-phase Euler equations modelling the problem of multi-phase hybrid detonations with many peaks in thermicity, a problem the authors gauge as being too complex to tackle head-on.  The full problem will be addressed in a sequel.

The paper is organized as follows. The mathematical model to be solved is given in section \ref{sec:Model}. The possible steady travelling wave solutions, akin to the Zeldovich-von Neumann-D\"{o}ring structure, are presented in section \ref{sec:Steady_analytical_solution}. Integral curves of the steady solution are studied, and a Hugoniot-Rayleigh analysis is performed in section \ref{sec:Steady_analytical_results}. Unsteady numerical simulations of the model are then presented; the method is described in section \ref{sec:Numerical_method} and the results are shown in \ref{sec:Numerical_results}. Solutions, stability, and selection rules are then discussed in section \ref{sec:Discussion} in light of the analytical and numerical results. The conclusions are summarized in section \ref{sec:Conclusion}.

\section{\label{sec:Model}Model: Reactive Burgers equation}

The formulation follows closely that of Fickett \cite{fickett1979detonation,fickett1985introduction}, along with the interpretation of the different terms.
The hydrodynamic model is
\begin{equation}
\label{eq:Hydrodynamic}
\partial_t \rho + \partial_x p = 0
\end{equation}
where $x$ is a Lagrangian coordinate, $t$ is time, $\rho$ represents density, and $p$ is the mass flux but serves as the analogue to pressure, by which it will be referred to in the remainder of this article. In this study, the equation of state
\begin{equation}
\label{eq:EOS}
p = \frac{1}{2} \left(\rho^2 + \sum_i \lambda_i Q_i\right)
\end{equation}
is used, where $\lambda_i$ are the reaction progress variables which range from zero (unreacted) to one (reacted). The constants $Q_i$ represent the heat release when positive, and losses when negative.

Two sequential exothermic reactions (subscripts 1 and 2) and one loss (subscript 3) are considered with state-independent rates. The first exothermic reaction and the heat loss begin once shocked, while the second reaction begins upon completion of the first reaction. Simple depletion reaction rates are chosen for the exothermic reactions while a constant rate is used for the loss
\begin{equation}
\label{eq:ReactionRates}
\begin{aligned}
r_1 =& \partial_t\lambda_1= \text{H}(1-\lambda_1) k_1(1-\lambda_1)^{\nu_1},\\
r_2 =& \partial_t\lambda_2= (1-\text{H}(1-\lambda_1))\text{H}(1-\lambda_2) k_2(1-\lambda_2)^{\nu_2},\text{ and }\\
r_3 =& \partial_t\lambda_3 = \text{H}(1-\lambda_3) k_3,
\end{aligned}
\end{equation}
where $k_i$ are scaling constants, and $\nu_i$ represent the reaction orders. The Heaviside function 
\begin{align*}
\text{H}(n) = \begin{cases} 
0 &\text{ if } n \le 0 \\
1 &\text{ if } n > 0 \end{cases} 
\end{align*}
is used to toggle the reactions.

Initial conditions ahead of the wave are uniform, taking the values $p_0 = \rho_0 = \lambda_{i,0} = 0$ for simplicity and without loss of generality.

The parameters used in this study are listed in table \ref{tab:parameters}.
The parameters $k_1$, $Q_1$, and $\nu_1$ correspond to the scaling constant, heat release, and reaction order of the first exothermic reaction, $k_2$, $Q_2$, and $\nu_2$ to the subsequent exothermic reaction, and $k_3$ and $Q_3$ are for the loss. They have been chosen so that sonic points are possible for reactions 1 and 2, but their exact values and scale are otherwise arbitrary. Note the selection of $ 0 < \nu < 1$ allows the reaction to finish in finite time. 
Three cases are presented in the table, corresponding to different detonation velocities $D_{\mathrm{A}}$ and $D_{\mathrm{B}}$ that will be discussed in the next section. Only the heat release of the second reaction, $Q_2$, is changed, and with it $D_{\mathrm{B}}$.
This model was presented by Radulescu {et al.} \cite{radulescu2013multiplicity} for the single CJ point case and Lau-Chapdelaine {et al.} \cite{lau-chadeplaine2014existence} for the two CJ point case.

\begin{table}
	\begin{center}
		\caption{\label{tab:parameters}Parameter sets used in study ($\dagger$ 1 CJ point, $\ddagger$ 2 CJ points)} %
		\begin{tabular}{l c || c c c | c c c | c c || c c }
			\toprule
			case & & $k_1$ & $Q_1$ & $\nu_1$ & $k_2$ & $Q_2$ & $\nu_2$ & $k_3$ & $Q_3$ & $D_{\mathrm{A}}$ & $D_{\mathrm{B}}$
			\\ \hline
			$D_{\mathrm{A}} < D_{\mathrm{B}}$ & $\dagger$ & 1 & 0.5 & 0.5 & 0.2 & 0.5 
			& 0.5 & 0.1 & -0.8 & 0.59397 & 0.6
			\\
			$D_{\mathrm{A}} {=} D_{\mathrm{B}} {=} D_{\mathrm{AB}}$ & $\ddagger$ & 1 & 0.5 & 0.5 & 0.2 & 0.4782... 
			& 0.5 & 0.1 & -0.8 & 0.59397 & 0.59397
			\\
			$D_{\mathrm{A}} > D_{\mathrm{B}}$ & $\dagger$ & 1 & 0.5 & 0.5 & 0.2 & 0.45 
			& 0.5 & 0.1 & -0.8 & 0.59397 & 0.58784
			\\ 
			\bottomrule
		\end{tabular}	
	\end{center}
\end{table}	

\section{\label{sec:Steady_analytical_solution}Steady analytical solution}
A travelling wave solution can be deduced when the hydrodynamic equation \ref{eq:Hydrodynamic} is written in characteristic form
\begin{align}
\label{eq:Characteristics}
\frac{\text{d}p}{\text{d} t} = \sigma \text{ along } \frac{\text{d} x}{\text{d} t} = \rho
\end{align}
where thermicity is defined as
\begin{align}
\label{eq:thermicity}
\sigma = \frac{1}{2} \sum_i r_i Q_i.
\end{align}

\begin{figure}
	\centering
	\includegraphics[scale=0.5]{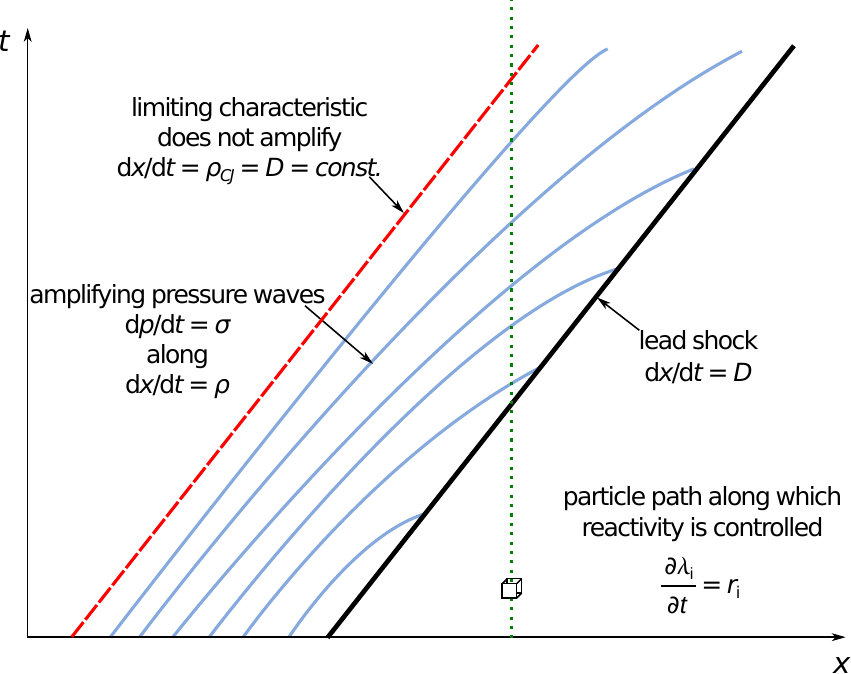}
	\caption{The structure of self-sustained detonations}
	\label{fig:Steady_state_detonation_structure}
\end{figure}

Figure \ref{fig:Steady_state_detonation_structure} illustrates the structure of the self-supported detonation wave \cite{radulescu2011nonlinear}. The detonation structure consists of pressure waves originating from the back (left), travelling along characteristics ${\text{d}x}/{\text{d}t} = \rho$ and amplifying according to the characteristic equation \eqref{eq:Characteristics}. The amplification is given by the local evolution of the reacting field, described by the reaction rate equations \eqref{eq:ReactionRates}, along the particle paths $x = \text{constant}$. These pressure waves coalesce and sustain a steady moving lead shock with velocity $D$.

In order to seek the steady structure of the travelling wave solution illustrated in figure \ref{fig:Steady_state_detonation_structure}, the spatial variable $x$ is changed to a shock fixed frame $\zeta = x - D t$ (with origin at the shock) for a detonation travelling with constant velocity $D$. The hydrodynamic equation \eqref{eq:Hydrodynamic} and reaction rate equations \eqref{eq:ReactionRates} can then be transformed into a set of ordinary differential equations
\begin{align}
\frac{\text{d} \rho}{\text{d} \zeta} &= \frac{1}{D} \frac{\sigma}{\rho-D}
\text{, } \label{eq:SS_ODE_density} \\
\frac{\text{d} \lambda_1}{\text{d} \zeta} &= -\frac{r_1}{D}
\text{, } \nonumber \\
\frac{\text{d} \lambda_2}{\text{d} \zeta} &= -\frac{r_2}{D}
\text{, and } \nonumber \\
\frac{\text{d} \lambda_3}{\text{d} \zeta} &= -\frac{r_3}{D}. \nonumber
\end{align}
Integrating these equations yields the analytical results for the wave structure
\begin{align}
\rho =& D \pm \sqrt{D^2 - \left(\sum_i Q_i \lambda_{i}\right)}, \label{eq:SSDensity}
\\ \lambda_1 =& 1- \left(-\frac{k_1}{D}(\nu_1-1) \zeta + 1\right)^{\frac{1}{1-\nu_1}}, \label{eq:SSReactionProgress1}
\\ \lambda_2 =& 1- \left(\frac{k_2}{D}(\nu_2-1)(\zeta_2 - \zeta) + 1\right)^{\frac{1}{1-\nu_2}}, \label{eq:SSReactionProgress2}
\\ \lambda_3 =& -\frac{k_3}{D} \zeta, \label{eq:SSReactionProgress3}
\end{align}
where $\zeta_2 = \frac{D}{k_1(\nu_1-1)}$ indicates where the second reaction starts, found by solving equation \ref{eq:SSReactionProgress1} for $\zeta$ with $\lambda_1 = 1$; the first and third reaction begin at $\zeta=0$.

The travelling wave solution is isolated from the back when the limiting characteristic travels at the same velocity as the steady lead shock, {i.e.}, when ${\text{d}x}/{\text{d}t} = \rho = D$; this is the sonic criterion. For this limiting characteristic to travel at constant speed, it also requires vanishing thermicity from the density differential equation \eqref{eq:SS_ODE_density}. Thus the generalized Chapman-Jouguet condition, denoted with the subscript CJ, is fulfilled when
\begin{align}
\rho = D && \text{ and } && \sigma = \frac{1}{2} \sum_i r_{i} Q_i = 0. \label{eq:Generalized_CJ_condition}
\end{align}

The solution exhibits sonic points, with respect to the front, anywhere the condition $\rho = D$ is met. The relation between the detonation speed and the reaction, evaluated along the limiting characteristic, is obtained from the analytical relation for density \eqref{eq:SSDensity} when the sonic portion of the generalized CJ condition \eqref{eq:Generalized_CJ_condition} is met, and is
\begin{align}
\label{eq:Detonation_velocity}
D^2 = \sum_i Q_i \lambda_{i}.
\end{align}
This expression illustrates that the detonation velocity for the solution with sonic points is given by the net energy evolved from the lead shock to the sonic plane.

The second portion of the generalized CJ condition \eqref{eq:Generalized_CJ_condition}, the balance of reaction rates (zero thermicity condition), permits the reaction progress values along the limiting characteristic to be established. Since the first and second exothermic reactions are sequential, the balance of rates must occur between either the first and loss (third) reaction, or the second reaction and the loss, {i.e.} $r_{1} Q_{1} + r_{3} Q_{3} = 0$ or $r_{2} Q_{2} + r_{3} Q_{3} = 0$. Denoting these sonic points as A and B respectively, two solutions are obtained
\begin{equation}
\begin{aligned}
\label{eq:CJ_A}
& \lambda_{1\mathrm{A}} = 1-\left(-\frac{k_3 Q_3}{k_1 Q_1}\right)^{\frac{1}{\nu_1}} 
\text{, } \lambda_{2\mathrm{A}} = 0 \text{, and }\\
& \lambda_{3\mathrm{A}} = \frac{k_3}{k_1 (\nu_1-1)}\left(\left(-\frac{k_3 Q_3}{k_1 Q_1}\right)^{\frac{1-\nu_1}{\nu_1}} -1 \right)
\end{aligned}
\end{equation}
for the sonic point closest to the shock, and
\begin{equation}
\begin{aligned}
\label{eq:CJ_B}
\lambda_{1\mathrm{B}} =& 1
\text{, } \lambda_{2\mathrm{B}} = 1-\left(-\frac{k_3 Q_3}{k_2 Q_2}\right)^{\frac{1}{\nu_2}}
\text{, and }\\ \lambda_{3\mathrm{B}} =& - \frac{k_3}{k_1(\nu_1-1)}+ \frac{k_3}{k_2(\nu_2-1)}\left(\left(-\frac{k_3 Q_3}{k_2 Q_2}\right)^{\frac{1-\nu_2}{\nu_2}}-1\right)
\end{aligned}
\end{equation}
for the second. 
The sonic point positions are found by substituting these results into the wave structure equations (\ref{eq:SSReactionProgress1}, \ref{eq:SSReactionProgress2}, and \ref{eq:SSReactionProgress3}),
\begin{align*}
&\zeta_{\mathrm{A}} = \zeta_1 - \frac{D}{k_1(\nu_1-1)}\left(\left(-\frac{k_3 Q_3}{k_1 Q_1}\right)^{\frac{1-\nu_1}{\nu_1}}-1\right)
\text{, and }\\ 
&\zeta_{\mathrm{B}} = \zeta_1 + \frac{D}{k_1(\nu_1-1)} - \frac{D}{k_2(\nu_2-1)}\left(\left(-\frac{k_3 Q_3}{k_2 Q_2}\right)^{\frac{1-\nu_2}{\nu_2}}-1\right).
\end{align*}
The solution is now complete for the detonation speed and reaction zone structure in closed algebraic form.

\section{\label{sec:Steady_analytical_results}Steady analytical results}

\subsection{Integral curves}

\begin{wrapfigure}{r}{0.5\textwidth}
	\centering
	\begin{subfigure}[t]{0.5\textwidth}
		\includegraphics[width=\textwidth]{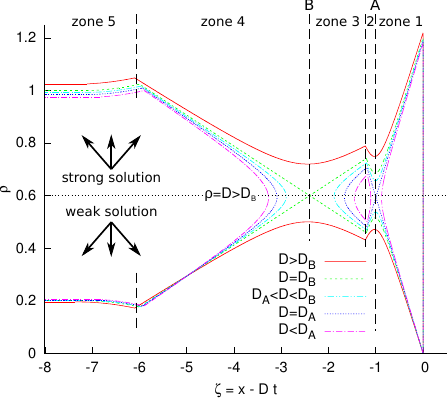}
		\caption{One CJ point, $D_{\mathrm{A}} = 0.59397 < D_{\mathrm{B}} = 0.6 $}
		\label{fig:Steady_state_1_CJ_point}
	\end{subfigure}%
	\\
	\begin{subfigure}[t]{0.5\textwidth}
		\includegraphics[width=\textwidth]{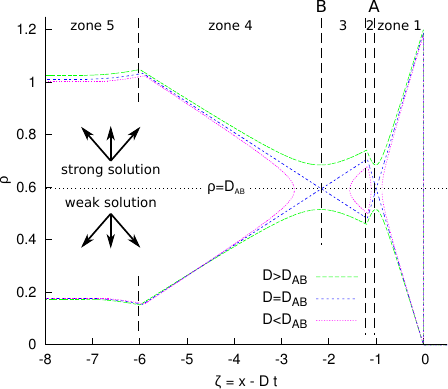}
		\caption{Two CJ points, $D_{\mathrm{A}} = D_{\mathrm{B}} = D_{\mathrm{AB}} = 0.59397 $}
		\label{fig:Steady_state_2_CJ_points}
	\end{subfigure}%
	\\
	\begin{subfigure}[t]{0.5\textwidth}
		\includegraphics[width=\textwidth]{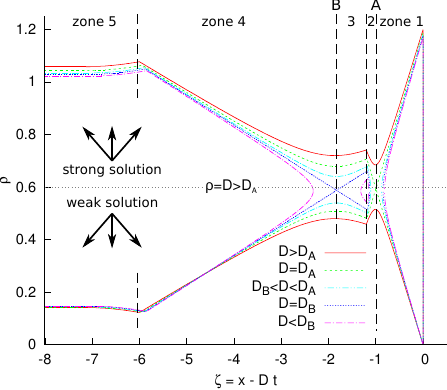}
		\caption{One CJ point, $D_{\mathrm{A}} = 0.59397 > D_{\mathrm{B}} = 0.58784 $}
		\label{fig:Steady_state_1_CJ_points,DA>DB}
	\end{subfigure}%
	\caption{Family of steady integral curves for a detonation travelling from left to right with parameters found in table \ref{tab:parameters}}
	\label{fig:Steady_state}
\end{wrapfigure}

To illustrate the type of solution obtained, consider an example with parameters $D_{\mathrm{A}} < D_{\mathrm{B}}$, where $D_{\text{A}}$ is the CJ detonation speed determined by sonic point A (equations \ref{eq:Detonation_velocity} and \ref{eq:CJ_A}), and $D_{\text{B}}$ for sonic point B (equations \ref{eq:Detonation_velocity} and \ref{eq:CJ_B}). Figure \ref{fig:Steady_state_1_CJ_point} shows four families of integral curves for the parameters listed in table \ref{tab:parameters} with $D_{\mathrm{A}} < D_{\mathrm{B}}$ and one CJ point. For these parameters, $D_{\mathrm{A}} = 0.59397$ and $D_{\mathrm{B}} = 0.6$, while the equilibrium detonation speed is $D_{\text{eq}} = \sqrt{Q_1 + Q_2 + Q_3} = 0.447$. The integral curves begin at the shock ($\zeta = 0$) with a value of $\rho$ given by the inert shock jump conditions in Burgers equation, $\rho = 2D$, and proceed towards the burned side.

A steady shock velocity $D$ is chosen, paying attention to its relation to $D_{\mathrm{B}}$ and $D_{\mathrm{A}}$. When $D > D_{\mathrm{B}}$, the integral curve (top most curve in figure \ref{fig:Steady_state_1_CJ_point}) does not intersect any sonic point $\rho=D$; this is the over-driven solution which requires that the rear boundary be maintained at the corresponding value. The evolution of $\rho$ is non-monotonous.
Initially (zone 1), the heat release is greater than the loss, {i.e.} ${r_1 Q_1 + r_3 Q_3 > 0}$, and the net positive thermicity leads to positive density and pressure gradients, owing to the amplification of forward-facing pressure waves. The first zero in density gradient corresponds to when $r_1 Q_1 + r_3 Q_3 = 0$ (point A), the first zero in thermicity. Towards the left (zone 2), $r_1 Q_1 + r_3 Q_3 < 0$, the density increases as pressure waves are attenuated. Once the second reaction begins and overcomes the endothermic processes (zone 3), $r_2 Q_2 + r_3 Q_3 > 0$, density once again decreases towards the left. After the second point of vanishing thermicity (point B), obtained when $r_2 Q_2 + r_3 Q_3 = 0$, the losses overcome the second exothermic reaction (zone 4). The last segment (zone 5) corresponds to when the losses terminate and the second reaction eventually comes to equilibrium.

For a detonation speed corresponding to the largest eigenvalue, $D = D_{\mathrm{B}}$ in this case, a sonic point appears at point B through the balance of the second reaction and the losses. Note that this sonic point is a saddle point (see section \ref{sec:Lin_stab_analysis}), and both weak and strong solutions can be attained on its left, depending on the rear boundary conditions. The strong solution is subsonic with respect to the detonation front and the weak solution is supersonic with respect to the detonation front. This means signals originating within the strong solution can catch up to the shock while signals from the weak solution cannot. This is the typical behaviour of pathological detonations, and well discussed by Fickett \cite{fickett1985introduction}.

Sonic point B disappears at detonation speeds between the eigenvalues, $D_{\mathrm{A}} < D < D_{\mathrm{B}}$. The integral curves corresponding to this solution terminate at a singularity in zone 3.
When $D = D_{\mathrm{A}}$, a sonic point occurs at point A, where the first reaction balances the losses.
However, a steady solution does not exist, owing to the singularity established in the reaction zone. It can thus be asserted that the smaller eigenvalue is not stable.

For detonation speeds lower than both eigenvalues, sonic point A disappears and the integral curves terminate at a singularity in zone 1.
The equilibrium solution, where the detonation velocity is determined by the total heat release, is thus not possible.

The velocities $D_{\mathrm{A}}$ and $D_{\mathrm{B}}$, calculated using equation \ref{eq:Detonation_velocity} at the CJ points, are determined by the parameter set listed in table \ref{tab:parameters}. Altering these parameters changes the values of $\lambda_i$ at the CJ points, through equations \ref{eq:CJ_A} and \ref{eq:CJ_B}, and therefore the detonation velocity associated with those CJ points. It is possible to reduce $D_{\mathrm{B}}$ by reducing the heat release of the second reaction, $Q_2$, for example.

Reducing $Q_2$ to the value in the second row of table \ref{tab:parameters} lowers the velocity associated with the second sonic point to be equal to the first sonic point velocity, $D_{\mathrm{A}} = D_{\mathrm{B}} = D_{\mathrm{AB}}$. The possibility of two sonic points appearing simultaneously in a steady solution arises. The steady shock speed determined by the generalized CJ criterion is equal for points A and B. This solution corresponds to integral curves passing through two saddle points, as shown in figure \ref{fig:Steady_state_2_CJ_points}. The weak and strong solutions are available to the left of both sonic points. Selection of the solution, strong or weak, between points A and B is independent of the rear boundary conditions because information from the rear boundary cannot cross sonic point B.
When the detonation is over-driven, $D>D_{\mathrm{AB}}$, the integral curve does not intersect the sonic points and must be supported by the rear boundary condition. A steady solution does not exist when $D<D_{\mathrm{AB}}$ as singularities are established near points A and B.

Further reducing the second heat release (e.g., third row of table \ref{tab:parameters}) lowers the second CJ point's detonation velocity below that of the first, $D_{\mathrm{B}} < D_{\mathrm{A}}$. In this case, the largest eigenvalue, $D_{\mathrm{A}}$ corresponds to the singularity free solution as shown in figure \ref{fig:Steady_state_1_CJ_points,DA>DB}. Over-driven $D > D_{\mathrm{A}}$ solutions exist, and singularities occur when $D < D_{\mathrm{A}}$. When $D = D_{\mathrm{B}} < D_{\mathrm{A}}$, a sonic point appears at B and a singularity is present at A.

\subsection{\label{sec:Hugoniot_and_Rayleigh_line_analysis}Hugoniot and Rayleigh line analysis}
The multiplicity of solutions can also be represented in $\rho$-$p$ phase space analysed by traditional Hugoniot-type arguments \cite{fickett1979detonation,fickett1985introduction}. Constructing a solution requires connecting possible unburned and burned loci (Hugoniots) with possible integral curves (Rayleigh lines) intersecting the loci of zero-thermicity ({e.g.} points A and B shown in figure \ref{fig:Steady_state_1_CJ_point}). Such a representation offers further insight into the solution.

\begin{figure*}[h!]
	\centering
	\includegraphics[width=\textwidth]{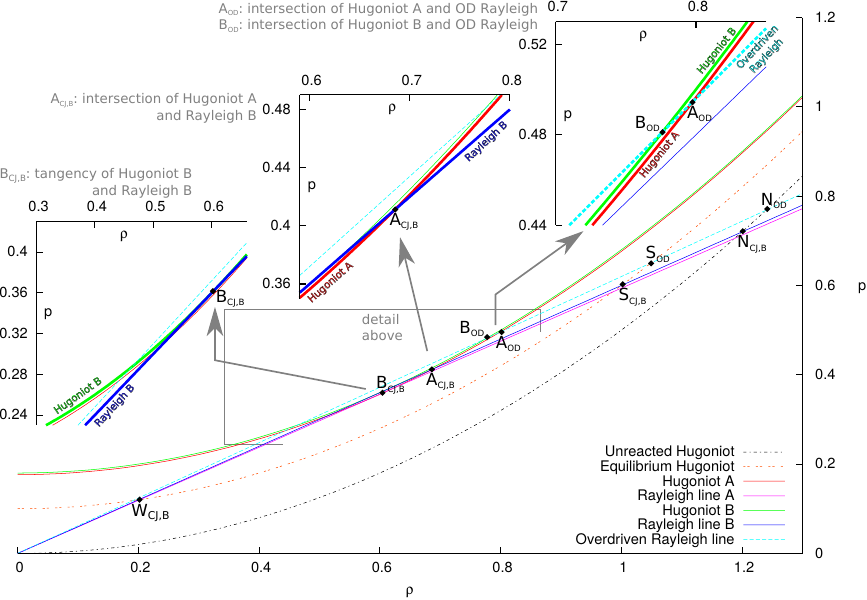}
	\caption{Rayleigh lines and Hugoniot curves on a $\rho$-$p$ diagram with the leading shock (N), equilibrium (S and W), and the two zero-thermicity (A and B) loci for the single steady sonic point properties; two sets of points are shown: an over-driven detonation ($D>D_{\mathrm{B}}=0.6>D_{\mathrm{A}} = 0.59397$, shown with subscript OD), and a detonation travelling at the CJ velocity corresponding to a sonic point at point B only ($D=D_{\mathrm{B}}>D_{\mathrm{A}}$, shown with subscript CJ,B); N represents the unreacted shock condition, S and W correspond to the strong and weak equilibrium solutions respectively}
	\label{fig:hugoniot}
\end{figure*}

For the simple Fickett model \cite{fickett1985introduction} considered here, the Hugoniots are simply given by the equation of state \ref{eq:EOS}. The unreacted shock Hugoniot ($\lambda_1 = \lambda_2 = \lambda_3 = 0$), the equilibrium Hugoniot ($\lambda_1 = \lambda_2 = \lambda_3 = 1$), the Hugoniot corresponding to the eigenvalue A ($\lambda_{1} = \lambda_{1\text{A}}$, $\lambda_{2} = \lambda_{2\text{A}}$, $\lambda_{3} = \lambda_{3\text{A}}$) and the Hugoniot corresponding to the eigenvalue B ($\lambda_{1} = \lambda_{1\text{B}}$, $\lambda_{2} = \lambda_{2\text{B}}$, $\lambda_{3} = \lambda_{3\text{B}}$), are shown in the $\rho$-$p$ phase space of figure \ref{fig:hugoniot} for the single sonic point example studied in the previous section, where $D_{\mathrm{A}} < D_{\mathrm{B}}$ and whose integral curves are shown in figure \ref{fig:Steady_state_1_CJ_point}. Rewriting equation \ref{eq:SS_ODE_density} as $\text{d}(p-\rho D)/\text{d}\zeta=0$, with boundary conditions at the leading shock ($p = 2D^2$, $\rho = 2D$, and labelled as N in figure \ref{fig:hugoniot}), they form the Rayleigh lines
\begin{align}
p = \rho D \label{eq:Rayleigh_line}.
\end{align}
which are shown in figure \ref{fig:hugoniot} for the eigenvalue B and the over-driven solution of figure \ref{fig:Steady_state_1_CJ_point}.

The over-driven integral curve ($D>D_{\mathrm{B}}$, subscript OD in figure \ref{fig:hugoniot}) starts at the leading shock, point N$_{\text{OD}}$, travels down the over-driven Rayleigh line, reaches point A$_{\text{OD}}$ on the eigenvalue Hugoniot A, and then B$_{\text{OD}}$ on the eigenvalue Hugoniot B, then returns to S$_{\text{OD}}$ on the strong branch of the equilibrium Hugoniot. Points A$_{\text{OD}}$ and B$_{\text{OD}}$ are the local minima in the matching integral curve of figure \ref{fig:Steady_state}, and correspond to the locus of zero thermicity, as explained above.

The integral curve of eigenvalue B ($D=D_{\mathrm{B}}$, subscript CJ,B) in figure \ref{fig:hugoniot} is tangent to Hugoniot B. It starts at point N$_\text{CJ,B}$, travels down the Rayleigh B line, intersects the Hugoniot A at point A$_\text{CJ,B}$ (first zero-thermicity point), then proceeds to the sonic point B$_\text{CJ,B}$ on the eigenvalue Hugoniot B, shown in detail at the top-left of figure \ref{fig:hugoniot}. Since this point is a saddle point, as shown in the integral curve, figure \ref{fig:Steady_state}, the solution can then reach either the strong solution S$_\text{CJ,B}$ or the weak solution W$_\text{CJ,B}$.

The eigenvalue A integral curve (not shown in figure \ref{fig:hugoniot}) is tangent to Hugoniot A and slightly lower than that of eigenvalue B for this case. Since Hugoniot A is lower than B, the corresponding velocity is also lower, seen by the Rayleigh line equation \ref{eq:Rayleigh_line}. Since this Rayleigh line never intersects Hugoniot B, a second zero thermicity point cannot be established in the system. The detonation speed selection rule and possible steady solution can thus also be made using the Hugoniot analysis. A regular solution requires intersection of the Rayleigh line with both zero-thermicity solutions. This can only be achieved for detonation velocities equal to, or larger than, the largest eigenvalue.

In the case where two simultaneous steady sonic points are possible, the Hugoniot curves A and B overlap.

\section{\label{sec:Numerical_method}Unsteady numerical method}
\begin{figure*}[t!]
	\centering
	\begin{subfigure}[t]{0.36\textwidth}
		\includegraphics[width=\textwidth]{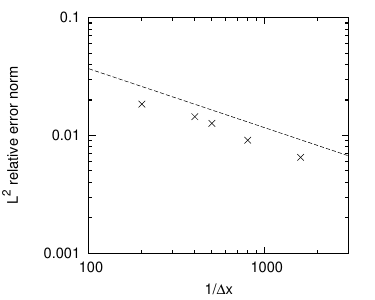}
		\caption{Effect of resolution on density profile error ($t=200$, $\text{CFL}=0.5$); dashed line shows $(1/\Delta x) ^{-1/2}$}
		\label{subfig:piston_L2_mean_error_norm}
	\end{subfigure}%
	\begin{subfigure}[t]{0.64\textwidth}
		\includegraphics[width=\textwidth]{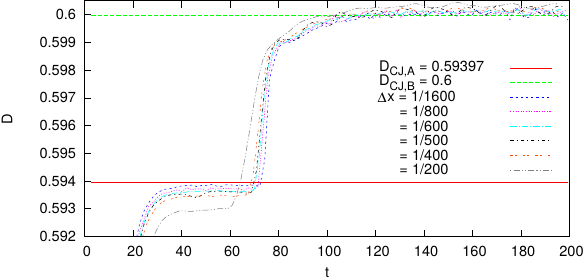}
		\caption{Effect of resolution on detonation speed ($\text{CFL}=0.5$)}
		\label{subfig:piston_shock_speed_resolution}
	\end{subfigure}%
	\\	\begin{subfigure}[t]{0.36\textwidth}
		\includegraphics[width=\textwidth]{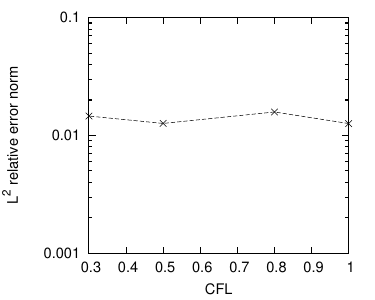}
		\caption{Effect of time step on density profile error ($t=200$, $\Delta x=1/500$)}
		\label{subfig:piston_CFL_L2_mean_error_norm}
	\end{subfigure}%
	\begin{subfigure}[t]{0.64\textwidth}
		\includegraphics[width=\textwidth]{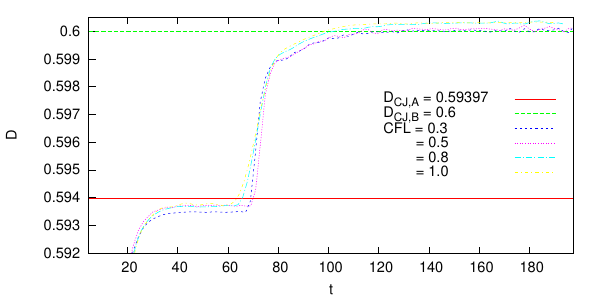}
		\caption{Effect of time step on detonation speed ($\Delta x=1/500$)}
		\label{subfig:piston_CFL_shock_speed_resolution}
	\end{subfigure}%
	\caption{Effect of resolution and time step size on piston-initiated detonation described in section \ref{sec:Single CJ point}}
	\label{fig:resolution_study}
\end{figure*}

Unsteady numerical simulations were used to study the stability and transients of the model. The domain was discretized with a uniform grid spacing of $\Delta x = 1/500$. 
A resolution study was performed, using the piston-initiated detonation described in section \ref{sec:Single CJ point}. The results in figure \ref{subfig:piston_L2_mean_error_norm} show that the error, calculated by the L$^2$ relative error norm 
$\sqrt{\sum(\rho_{\mathrm{exact}} - \rho_{\mathrm{numeric}})^2/\sum \rho_{\mathrm{exact}}^2}$
once the solution reached steady state, converges with increased resolution at the expected rate, and the quasi-steady-state detonation velocity approaches the analytic result, shown by figure \ref{subfig:piston_shock_speed_resolution}.

Time step size was determined using the Courant-Friedrichs-Lewy (CFL) condition such that $\text{CFL} =0.5 = \Delta x /( \Delta t \times \max(\rho))$, where $\max(\rho)$ is the maximum value of $\rho$ in the entire domain. A reaction threshold $\rho \ge 0.01$ was set to prevent unshocked gas from reacting. The effect of time step size, controlled by the $\text{CFL}$ number, is shown in figure \ref{fig:resolution_study}. While the steady state profile error increases slightly with decreasing time step, the detonation velocity approaches the CJ velocity and stabilizes at $\text{CFL}=0.5$ at a velocity slightly larger than CJ, seen in figure \ref{subfig:piston_CFL_shock_speed_resolution}.

The Riemann problem was solved at every cell interface using a first-order Godunov method as described by Clarke {et al.} \cite{clarke1989numerical}. Time evolution was done using a first-order upwind method.

A domain with an adaptive size was used where cells were dynamically added ahead of the shock, while cells in the rear were removed once the flow had equilibrated.

\section{\label{sec:Numerical_results}Unsteady numerical results}

\subsection{\label{sec:Single CJ point}Piston-initiated detonation}

\begin{figure*}[t!]
	\centering
	\begin{subfigure}[]{0.33\textwidth}
		\includegraphics[width=\textwidth]{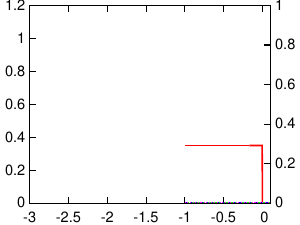}
		\caption{$t=0$}	\label{fig:piston_wave_profiles_piston:t=0}
	\end{subfigure}%
	\begin{subfigure}[]{0.33\textwidth}
		\includegraphics[width=\textwidth]{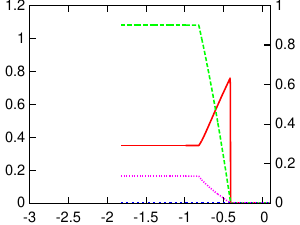}
		\caption{$t=1.37$}
		\label{fig:piston_wave_profiles_piston:t=1.36888}
	\end{subfigure}%
	\begin{subfigure}[]{0.33\textwidth}
		\includegraphics[width=\textwidth]{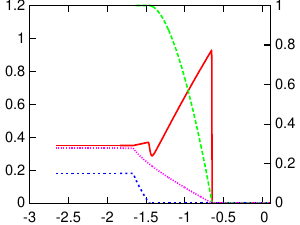}
		\caption{$t=2.78$}
		\label{fig:piston_wave_profiles_piston:t=2.77779}
	\end{subfigure}%
	\\	\begin{subfigure}[]{0.49\textwidth}
		\includegraphics[width=\textwidth]{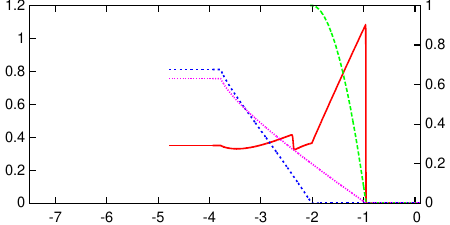}
		\caption{$t=6.31$}
		\label{fig:piston_wave_profiles_piston:t=6.3088}
	\end{subfigure} \hfill%
	\begin{subfigure}[]{0.49\textwidth}
		\includegraphics[width=\textwidth]{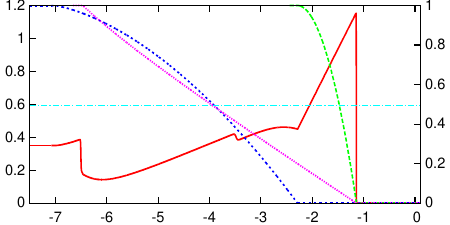}
		\caption{$t=11.6$}
		\label{fig:piston_wave_profiles_piston:t=11.6442}
	\end{subfigure}%
	\\	\begin{subfigure}[]{0.49\textwidth}
		\includegraphics[width=\textwidth]{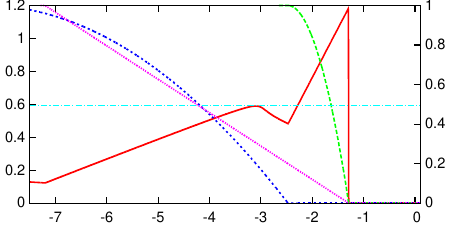}
		\caption{$t=25.6$}
		\label{fig:piston_wave_profiles_piston:t=25.6311}
	\end{subfigure} \hfill%
	\begin{subfigure}[]{0.49\textwidth}
		\includegraphics[width=\textwidth]{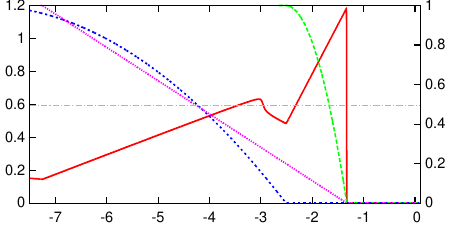}
		\caption{$t=30.0$}
		\label{fig:piston_wave_profiles_piston:t=30.0358}
	\end{subfigure}%
	\\	\begin{subfigure}[]{0.49\textwidth}
		\includegraphics[width=\textwidth]{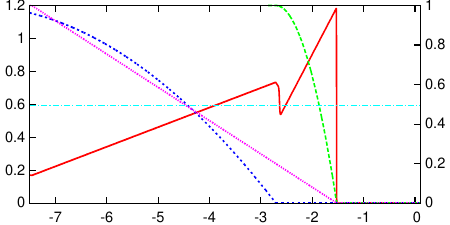}
		\caption{$t=61.5$}
		\label{fig:piston_wave_profiles_piston:t=61.5262}
	\end{subfigure} \hfill%
	\begin{subfigure}[]{0.49\textwidth}
		\includegraphics[width=\textwidth]{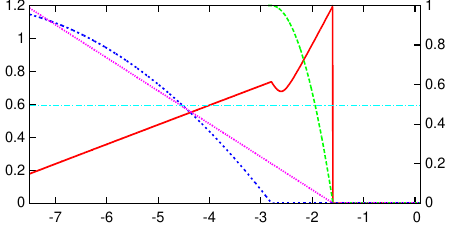}
		\caption{$t=80.1$}
		\label{fig:piston_wave_profiles_piston:end}
	\end{subfigure}%
	\\	\begin{subfigure}[]{1\textwidth}
		\centering
		\includegraphics[scale=1]{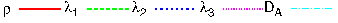}
		\caption{Legend: left axis for $\rho$, right axis for $\lambda_i$, horizontal axis for $\zeta=x-D_{\text{B}}t$}
		\label{fig:piston_wave_profiles_piston:legend}
	\end{subfigure}%
	\caption{Wave profiles for piston-initiated detonation for the single sonic point case where $D_{\text{CJ}} = D_{\text{B}} = 0.6 > D_{\text{A}} = 0.59397$}
	\label{fig:wave_profiles_piston}
\end{figure*}

\begin{figure*}[]
	\centering
	\begin{subfigure}[b]{0.49\textwidth}
		\includegraphics[scale=0.83]{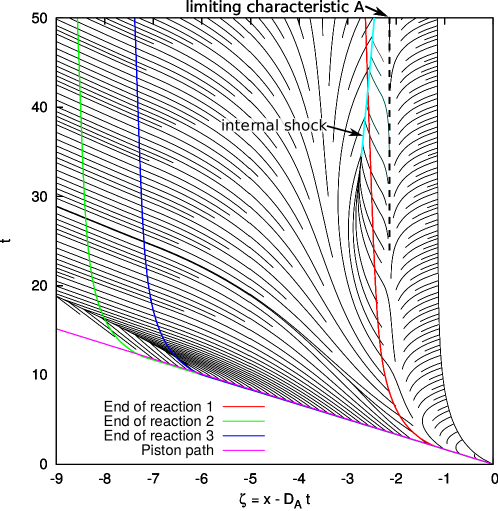}
		\caption{At early times}
		\label{subfig:piston_characteristicsA}
	\end{subfigure} \hfill%
	\begin{subfigure}[b]{0.49\textwidth}
		\includegraphics[scale=0.83]{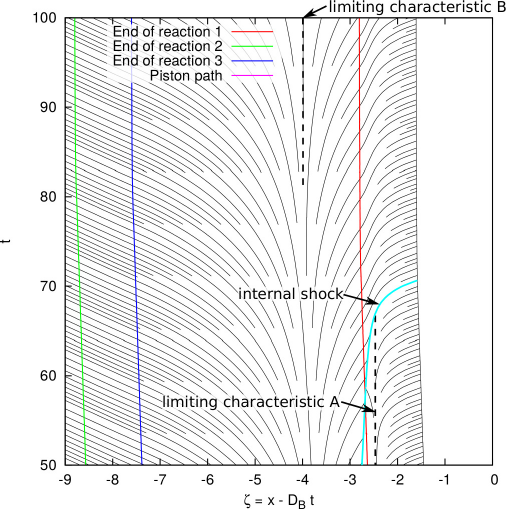}
		\caption{At later times}
		\label{subfig:piston_characteristicsB}
	\end{subfigure}%
	\caption{Characteristic diagram for a piston-initiated detonation (single sonic point case, $D_{\text{A}} =0.59397$, $D_{\text{B}} =0.6$)}
	\label{fig:piston_characteristics}
\end{figure*}

An example of the unsteady evolution of piston-initiated detonations, for conditions where the second sonic surface is slightly faster than the first ({i.e.} $D_{\text{B}} > D_{\text{A}}$), is shown with snapshots of the $\rho$ profile in figure \ref{fig:wave_profiles_piston} and with characteristics in figure \ref{fig:piston_characteristics}.

A piston with velocity $\rho=0.35$ was chosen in order to have an unsupported detonation, yet terminate at a value slightly larger than the weak solution. The initial condition can be seen in figure \ref{fig:piston_wave_profiles_piston:t=0}.

The first reaction initially drives a shock (figure \ref{fig:piston_wave_profiles_piston:t=1.36888}). An over-expansion occurs once the first reaction weakens, followed by a recompression at the start of the second reaction, seen in figure \ref{fig:piston_wave_profiles_piston:t=2.77779}. The recompression forms into a shock (figure \ref{fig:piston_wave_profiles_piston:t=6.3088}) which weakens and falls behind as the front continues to accelerate (figure \ref{fig:piston_wave_profiles_piston:t=11.6442}), and a compression wave is formed at the piston as the loss reaction terminates just before the end of the second heat release. The piston is now behind the end of the reaction zone. The variable $\rho$ increases at the beginning of the second reaction zone until it reaches the sonic condition, seen in figure \ref{fig:piston_wave_profiles_piston:t=25.6311}. The detonation front now travels at a speed $D_{\text{A}}$ corresponding to the sonic surface A. 

Recall that the $D = D_{\mathrm{A}}$ steady solution from section \ref{sec:Steady_analytical_results} terminates in a singularity behind sonic point A. In the context of this unsteady solution, a shock that travels faster than sonic surface A is created at sonic surface B (figure \ref{fig:piston_wave_profiles_piston:t=30.0358}), eventually penetrating into the first reaction zone. It can thus be asserted that the smaller eigenvalue $D_{\mathrm{A}}$ is not stable, but can be established as an intermediate transient. The shock wave that is formed behind the singularity eventually catches up to the lead shock. A truly steady detonation is eventually established, travelling with a speed corresponding to the largest predicted eigenvalue $D_{\text{B}}$, and with a new corresponding sonic surface B (figure \ref{fig:piston_wave_profiles_piston:end}).

The process can also be visualized with the characteristic diagrams of figure \ref{fig:piston_characteristics}. Characteristics outside the detonation's domain of influence ({i.e.} to the right of the front) are omitted. Recall that the model only admits the forward-facing family of characteristics (in the absolute frame of reference). This family of characteristics is shown in the frame of reference travelling at the velocity $D_{\text{A}}$ in figure \ref{subfig:piston_characteristicsA} and $D_{\text{B}}$ in figure \ref{subfig:piston_characteristicsB}.

Figure \ref{subfig:piston_characteristicsA} shows that the detonation front (the right-most characteristic) quickly becomes vertical as it stabilizes to a speed $D_{\text{A}}$. The limiting characteristic A forms at sonic point A, whose neighbouring characteristics diverge on either side. Characteristics to the left of the sonic point turn back towards sonic point A and create an internal shock where characteristics of the same family intersect, resulting in a two-valued discontinuity. Sonic point B appears at the same time with diverging characteristics on either side.

In the right frame of figure \ref{fig:piston_characteristics}, the internal shock is seen travelling towards the front, eventually overtaking the limiting characteristic A and reaching the detonation front. The front then accelerates to the velocity $D_{\text{B}}$. A limiting characteristic is then established at point B, parallel to the detonation front in the $\zeta$ -- $t$ plane.

The piston-initiated detonation travelled at the velocity $D_{\text{A}}$ once the generalized CJ criterion was satisfied at sonic point A. Figure \ref{fig:resolution_study} shows the detonation front velocity travelled steadily until the arrival of the internal shock at the front, after which it travelled steadily at the velocity $D_{\text{B}}$. The creation of the internal shock appears to coincide with appearance of sonic point B, but the reason for its formation may lie hidden behind the complex transient evolution of the problem.

\subsection{\label{sec:Double CJ point}Two simultaneous CJ points}
A small modification was made in the second heat release as discussed in section \ref{sec:Steady_analytical_results}, listed in the second row of table \ref{tab:parameters}. This allowed for two steady sonic points to exist simultaneously, both with the steady detonation velocity of point A. The possible integral curves for this example connecting the quiescent gas to rear conditions are shown in figure \ref{fig:Steady_state_2_CJ_points} when $D=D_{\mathrm{AB}}$. 

The piston-initiated detonation of the double sonic point solution, with the same piston velocity as the previous case, evolved qualitatively similar to the single sonic point case discussed in the previous section. Upon establishing the two sonic points A and B predicted by the analytical solution, a shock was produced at point B as before. The shock reached the front and a steady state emerged with only a single sonic point, as opposed to the two predicted by the integral curves. This discrepancy is reviewed in section \ref{sec:Discussion}.

The piston-initiated detonation reached the predicted steady state before creating a shock, meaning that analytical steady-state solutions with two simultaneous sonic points can be used as initial conditions to study the origin of the internal shock.

The steady state starts ($\zeta=0$) in the strong regime since the detonation is lead by a shock. Two options are available to the flow at each sonic point, following either the strong or weak solution, making for four possible steady states when two CJ points are present. This number is increased by the possibility of shocks joining the weak and strong solutions. The shock speed $S$ in the Burgers equation, simply given by the average of left and right states \cite{fickett1979detonation}
\begin{align}
S = \frac{1}{2} (\rho_{\text{left}} + \rho_{\text{right}}),
\end{align}
allows shocks that travel at exactly the leading shock speed. Internal shocks can only exist to the left of the sonic points where there can be a jump from the weak to the strong solution

The shock-free initial steady-state curves will be referred to with an abbreviated notation. This notation will break the flow profile into three sections: the rear solution, the solution between the sonic points, and the solution behind the leading shock. A strong solution will be abbreviated with the letter S and a weak solution with the letter W. For example, WWS would refer to a detonation whose rear boundary follows the weak solution, has the weak solution between sonic points, and necessarily has the strong solution behind the leading shock. The WWS initial condition can be seen in figure \ref{fig:wave_profiles_--+:t=0}.

Simulations initiated with internal shock waves, such as the one produced at point B, behaved in the same manner as those which evolved shocks on their own and are therefore excluded from further presentation.

\subsubsection{WWS initial condition}
\begin{figure}[t!]
	\centering
	\begin{subfigure}[t]{0.49\textwidth}
		\includegraphics[width=\textwidth]{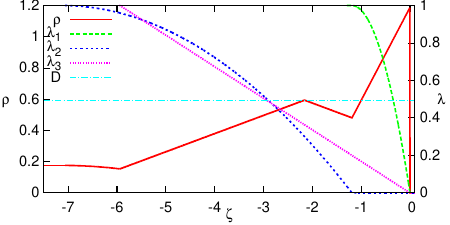}
		\caption{Initial condition ($t=0$)}	\label{fig:wave_profiles_--+:t=0}
	\end{subfigure} \hfill%
	\begin{subfigure}[t]{0.49\textwidth}
		\includegraphics[width=\textwidth]{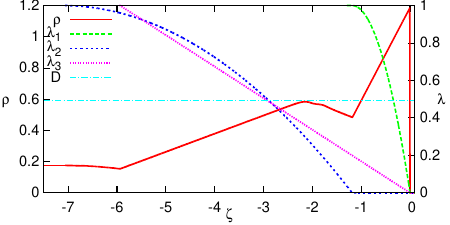}
		\caption{A disturbance approaches point B ($t=12$)}
		\label{fig:wave_profiles_--+:t=11.8494}
	\end{subfigure}%
	\\   	\begin{subfigure}[t]{0.49\textwidth}
		\includegraphics[width=\textwidth]{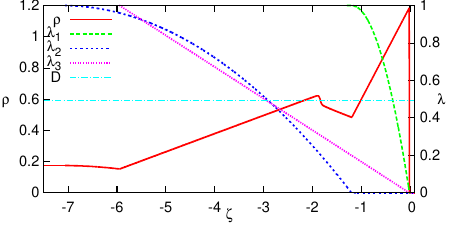}
		\caption{A shock forms and travels forward once the disturbance reaches point B ($t=44$)}
		\label{fig:wave_profiles_--+:t=44.0282}
	\end{subfigure} \hfill%
	\begin{subfigure}[t]{0.49\textwidth}
		\includegraphics[width=\textwidth]{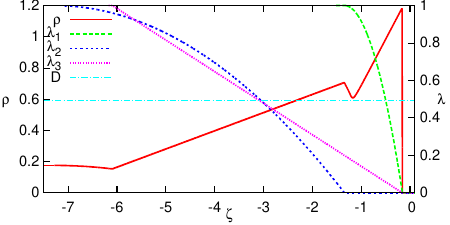}
		\caption{Detonation is slightly over-driven following the internal shock's arrival at the front ($t=450$)}
		\label{fig:wave_profiles_--+:end}
	\end{subfigure}%
	\caption{Wave profiles for WWS steady-state-initiated detonation for the double sonic point case ($D_{\text{CJ}}=D_{\text{AB}}=0.59397$)}
	\label{fig:wave_profiles_--+}
\end{figure}

The first simulation was initiated with the WWS integral curve described in the previous paragraph and seen in figure \ref{fig:wave_profiles_--+:t=0}. The exact WWS solution was imposed on the domain at $t = 0$ then evolved
through time. This solution can be achieved naturally with an under-driven piston, as demonstrated by the single CJ point, piston-initiated detonation seen in figure \ref{fig:piston_wave_profiles_piston:t=25.6311}, and behaves much in the same way.
Following initiation, a disturbance travels rearwards (figure \ref{fig:wave_profiles_--+:t=11.8494}) from the valley at the boundary between zones 2 and 3. A shock forms at point B following the arrival of the disturbance (figure \ref{fig:wave_profiles_--+:t=44.0282}). The shock travels faster than sonic point A, and eventually catches up to it and the front. During this transit, the detonation is slightly under-driven, while the final result is slightly over-driven and has a single sonic point at B (figure \ref{fig:wave_profiles_--+:end}), appearing qualitatively similar to case $D=D_{\mathrm{B}}$ in figure \ref{fig:Steady_state_1_CJ_point}.

\begin{figure*}[h!]
	\centering
	\includegraphics[width=\textwidth]{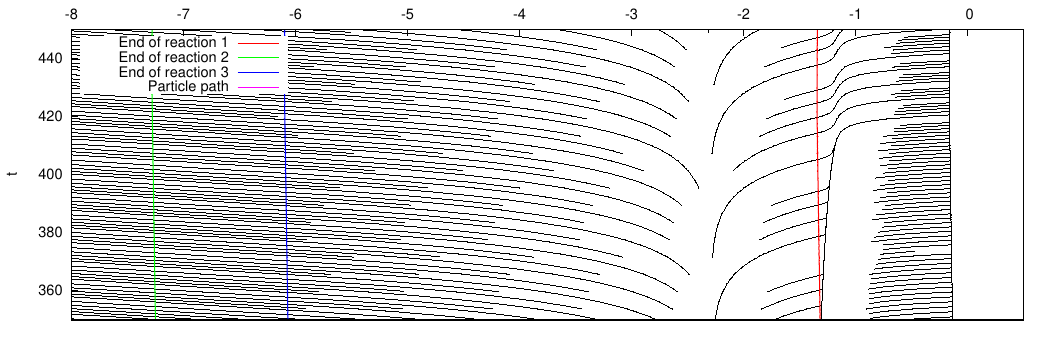}
	\\	
	\includegraphics[width=\textwidth]{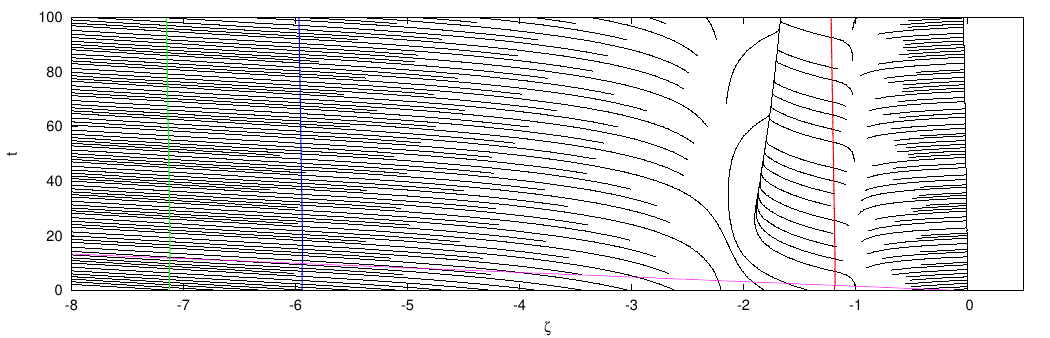}
	\caption{Characteristic diagram for the double sonic point detonation initiated with the WWS steady-state solution shown in figure \ref{fig:wave_profiles_--+}; intermediate time omitted in favour of clarity ($D_{\text{CJ}}=D_{\text{AB}}=0.59397$)}
	\label{fig:characteristics_--+}
\end{figure*}

The unsteady simulation can also be visualized using the characteristic diagram in figure \ref{fig:characteristics_--+}. Two quasi-steady runs of diverging characteristics are initially seen at points A and B.  The internal shock forms early on ($t<50$) and reaches the detonation front between $t=400$ and $t=450$. The detonation front is initially slightly under-driven and slopes slightly towards the left, but then slopes to the right as it becomes over-driven when the internal shock reaches the front.

\subsubsection{SWS initial condition}
\begin{figure}[h!]
	\centering
	\begin{minipage}{0.5\textwidth}%
		\begin{subfigure}[t]{0.98\textwidth}
			\includegraphics[width=\textwidth]{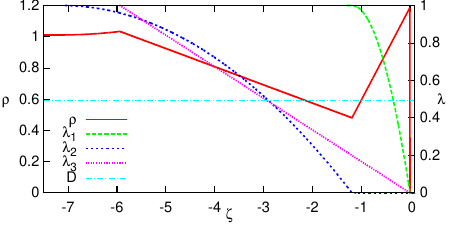}
			\vspace{-1.5\baselineskip}
			\caption{Initial condition ($t=0$)} \label{fig:wave_profiles_+-+:t=0}
		\end{subfigure}%
		\\
		\begin{subfigure}[t]{1\textwidth}
			\includegraphics[width=0.98\textwidth]{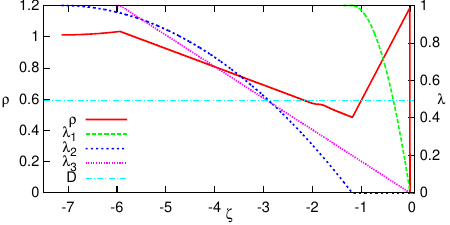}
			\vspace{-0.5\baselineskip}
			\caption{A disturbance approaches point B ($t=12$)}
			\label{fig:wave_profiles_+-+:t=11.8494}
		\end{subfigure}%
		\\
		\begin{subfigure}[t]{1\textwidth}
			\includegraphics[width=0.98\textwidth]{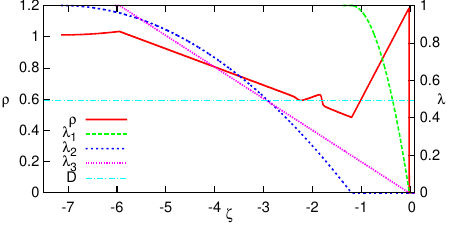}
			\vspace{-0.5\baselineskip}
			\caption{An internal shock forms and travels forward once the disturbance reaches point B ($t=52$)}
			\label{fig:wave_profiles_+-+:t=51.6496}
		\end{subfigure}%
		\\
		\begin{subfigure}[t]{1\textwidth}
			\includegraphics[width=0.98\textwidth]{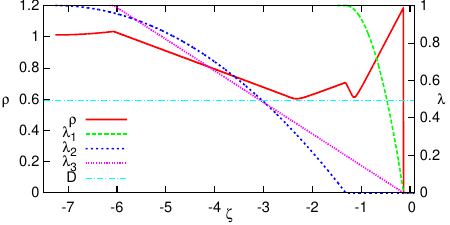}
			\vspace{-0.5\baselineskip}
			\caption{Points A and B are temporarily over-driven following the internal shock's arrival at the front ($t=351$)}
			\label{fig:wave_profiles_+-+:t=350.576}
		\end{subfigure}%
		\\
		\begin{subfigure}[t]{1\textwidth}
			\includegraphics[width=0.98\textwidth]{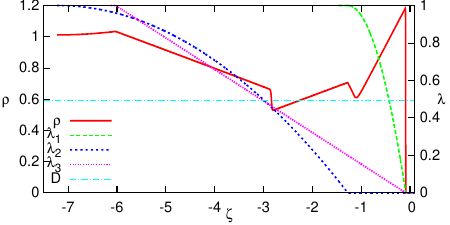}
			\vspace{-0.5\baselineskip}
			\caption{Point B becomes sonic and a backwards-travelling internal shock is formed ($t=600$)}
			\label{fig:wave_profiles_+-+:end}
		\end{subfigure}%
	\end{minipage}%
	\begin{minipage}{0.5\textwidth}%
		\begin{subfigure}[b]{1\textwidth}
			\centering
			\begin{minipage}{\textwidth}%
				\includegraphics[scale=1]{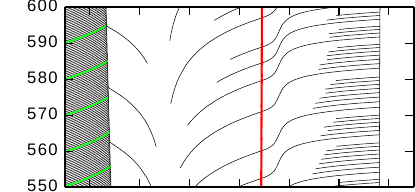} \\
				\includegraphics[scale=1]{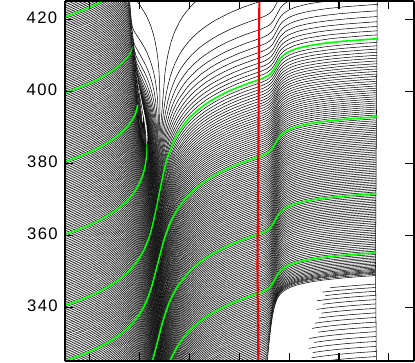} \\
				\includegraphics[scale=1]{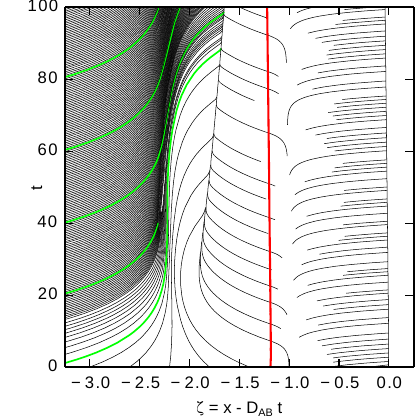}
			\end{minipage}
		\caption{\label{fig:characteristics_+-+}Characteristic diagram; intermediate times omitted, characteristics are periodically highlighted green, and the vertical red line indicates end of first reaction}
		\end{subfigure}\hfill%
	\end{minipage}
	\caption{\label{fig:SWS}Wave profiles and characteristic diagram for SWS steady-state-initiated detonation with the double sonic point case ($D_{\mathrm{CJ}}=D_{\text{AB}}=0.59397$)}
\end{figure}

Next consider a simulation initiated with the SWS integral curve shown in figure \ref{fig:wave_profiles_+-+:t=0}.
A disturbance (figure \ref{fig:wave_profiles_+-+:t=11.8494}) creates an internal shock (figure \ref{fig:wave_profiles_+-+:t=51.6496})
which catches up to the detonation front (figure \ref{fig:wave_profiles_+-+:t=350.576}) in the same manner as described in the WWS case. 
Another shock is then created downstream of sonic point B, joining newly supersonic flow (with respect to the detonation front) behind point B to subsonic flow of the rear boundary condition. This new shock wave distances itself from the front (figure \ref{fig:wave_profiles_+-+:end}).

The characteristics are shown in figure \ref{fig:characteristics_+-+}. Formation of the first internal shock occurs like in the WWS case. The second shock then forms behind sonic point B, where characteristics now diverge, and slowly travels away from the front. The rear of the domain is largely uniform with characteristics travelling towards the shock.

\subsubsection{SSS and WSS initial conditions}
The third and fourth possibilities, the SSS and WSS initial conditions, are not shown due to their simplicity. The SSS initial condition is nearly achieved in figure \ref{fig:wave_profiles_+-+:t=350.576} and evolves the same way to a slightly over-driven detonation with a shock behind the second sonic point, like figure \ref{fig:wave_profiles_+-+:end}.

For the WSS initial condition, point A drifts up into the strong solution, and the solution travels in a manner similar to the final state of figure \ref{fig:wave_profiles_--+:end}, again slightly over-driven.

\subsection{\label{sec:viscosity}Inclusion of viscosity}

In the case of two simultaneous steady CJ points, the velocity of both sonic points should be equal $D_{\mathrm{A}} = D_{\mathrm{B}} = D_{\mathrm{AB}}$ and the integral curves predict steady sonic points at A and B. Peculiarly, simulations of all initial conditions lead to an over-driven point A in the strong solution, instead of being sonic.

A hint to the reason can be found in the resolution study shown in figure \ref{subfig:piston_shock_speed_resolution}.
Low resolution simulations tend to travel slower than predicted by the generalized CJ criterion when sonic point A controls the detonation speed, and faster than predicted when the detonation speed is controlled by sonic point B. This discrepancy is reduced by increasing resolution but not eliminated.

While the inviscid analysis predicts that the sonic point speeds should be equal, $D_{\mathrm{A}} = D_{\mathrm{B}} = D_{\mathrm{AB}}$, numerical diffusion effectively causes the first sonic surface to travel slower than the second, $D_{\mathrm{A}} < D_{\mathrm{B}}$. When two sonic points are present, a perturbation from point B forces point A to become subsonic and the detonation adopts the faster velocity, as seen in the case with single steady CJ point, figure \ref{fig:piston_wave_profiles_piston:t=25.6311} onwards.

A `physical' viscosity was added to Fickett's detonation analogue hydrodynamic equation \ref{eq:Hydrodynamic} as
\begin{align}
\label{eq:Hydrodynamic_viscous}
\partial_t \rho + \partial_x p = \mu \partial_{xx} \rho
\end{align}
to assess whether this is a numerical artifact or if physical diffusion will have the same effect. The constant viscosity coefficient $\mu= 0.01$ was chosen to be an order of magnitude smaller than the heat release coefficients. The second-order derivative was discretized using centred differences, and the maximum time step size was selected by
\begin{align*}
\Delta t = \min \left(\mathrm{CFL} \times \frac{\Delta x}{\rho_{\mathrm{max}}}, \mathrm{CFL} \times \frac{1}{2} \frac{\Delta x^2}{\mu}\right)
\end{align*}
for numerical stability. The reaction threshold was raised to $\rho \ge 0.05$ to prevent reactions from starting in the diffusive foot of the incident shock. 
The viscous simulations with two simultaneous CJ points (section \ref{sec:Double CJ point}) were initiated with the ideal inviscid steady state profiles.

\begin{figure}[h!]
	\centering
	\begin{minipage}{0.5\textwidth}%
		\begin{subfigure}[t]{0.98\textwidth}
			\includegraphics[width=\textwidth]{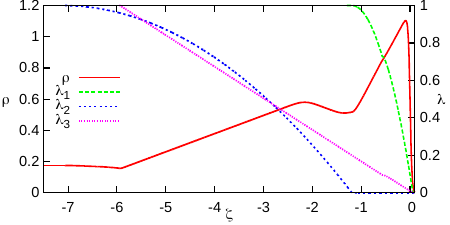}
			\vspace{-1.5\baselineskip}
			\caption{Rounding of corners ($t=1$)} \label{fig:wave_profiles_WWS:viscous:t=1}
		\end{subfigure}%
		\\
		\begin{subfigure}[t]{0.98\textwidth}
			\includegraphics[width=\textwidth]{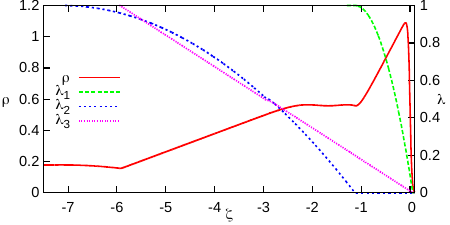}
			\vspace{-1.5\baselineskip}
			\caption{Flow between points A and B transitions from the weak to the strong solution ($t=4$)} \label{fig:wave_profiles_WWS:viscous:t=4}
		\end{subfigure}%
		\\
		\begin{subfigure}[t]{0.98\textwidth}
			\includegraphics[width=\textwidth]{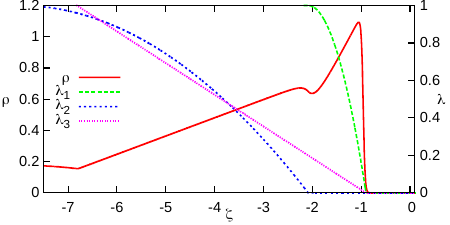}
			\vspace{-1.5\baselineskip}
			\caption{Steady viscous solution ($t=400$)} \label{fig:wave_profiles_WWS:viscous:t=400}
		\end{subfigure}%
	\end{minipage}%
	\begin{minipage}{0.5\textwidth}%
		\begin{subfigure}[b]{1\textwidth}
			\centering
		\includegraphics[width=0.8\textwidth]{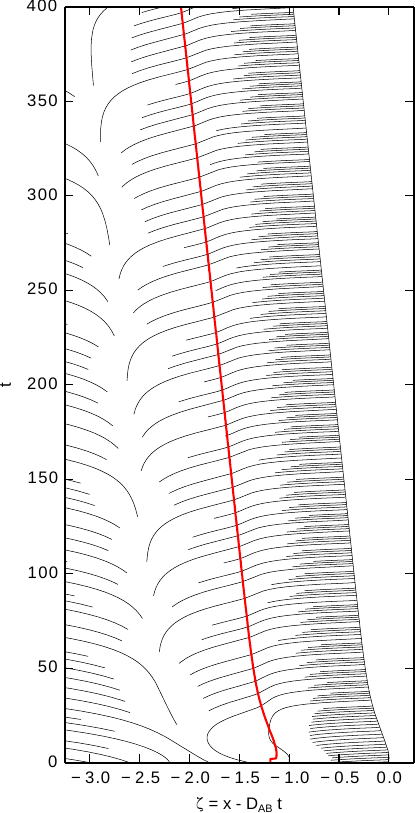}
			\caption{\label{fig:characteristics_WWS:viscous}Characteristic diagram; the vertical red line indicates end of first reaction}
		\end{subfigure}\hfill%
	\end{minipage}
	\caption{\label{fig:WWS:viscous}Wave profiles and characteristic diagram for the viscous WWS initiated with the inviscid steady-state detonation profile (figure \ref{fig:wave_profiles_--+:t=0}) with the double sonic point case ($\mu = 0.01$, $D_{\text{AB}}=0.59397$)}
\end{figure}

In the WWS case, the sharp corners of the initial profile are quickly rounded (figure \ref{fig:wave_profiles_WWS:viscous:t=1}) and the shock slows, as seen by the negative slope of the shock path in figure \ref{fig:characteristics_WWS:viscous}. Signals near point B now travel faster than the decayed shock and begin to move towards the front. The valley between the points A and B transitions (figure \ref{fig:wave_profiles_WWS:viscous:t=4}) from the weak solution to the strong solution and point A becomes non-sonic. The shock accelerates to a steady speed $D<D_{\mathrm{AB}}$, seen by the change in slope of the shock path in figure \ref{fig:characteristics_WWS:viscous}, driven by the net heat release between the shock and sonic point B. The WWS initial condition has transitioned to a WSS detonation with a non-sonic point A (figure \ref{fig:wave_profiles_WWS:viscous:t=400}), like the inviscid case in figure \ref{fig:wave_profiles_--+:end} but travelling slower due to the additional viscous losses.

Detonations initiated with the SWS or SSS cases end up qualitatively resembling the $D>D_{\mathrm{AB}}$ case of figure \ref{fig:Steady_state_2_CJ_points}, without sonic points in the solution as both points A and B become overdriven. A steady, overdriven viscous SSS solution is established, supported by the rear boundary condition at the inviscid CJ speed.

The addition of physical viscosity exaggerates the overdrive of point A seen in the inviscid two CJ point cases. Viscosity slows the detonation below its inviscid CJ speed, because of the additional viscous losses. The state-dependent diffusive losses are more important between the front and point A than point B, under these parameters, causing point A to become overdriven when two CJ points would coexist without viscosity. 

This suggests the discrepancy seen in the inviscid simulations is caused by numerical diffusive losses, since a similar structure is observed in viscous calculations. Numerical error causes the detonation to travel slightly faster than CJ and this difference is communicated to the front through the creation of an internal shock at point B. 

\section{\label{sec:Discussion}Creation of internal shocks}

The piston initiation of a detonation with a single steady CJ point yielded the steady integral curves. As the detonation front accelerated from the piston face, it began with a velocity $D<D_{\text{A}}<D_{\text{B}}$. The detonation velocity temporarily stabilized when $D=D_{\text{A}}$ because the creation of sonic point A stopped signals from the rear boundary from affecting the front. A shock formed behind point A, as inferred by the singularity in the integral curve (figure \ref{fig:Steady_state_1_CJ_point}), and sonic point B appeared behind the shock. Both sonic points A and B coexisted until the internal shock reached the detonation front, accelerating it to $D=D_{\text{B}}$. The detonation then continued steadily with the single sonic point predicted in the integral curves of figure \ref{fig:Steady_state_1_CJ_point}.

This supports the results of unsteady simulations by Khasainov and Veyssi\`{e}re \cite{khasainov1996initiation} that found the regime with a slower detonation speed was always accelerated to the faster velocity regime when multiple solutions were possible.

The same sequence of events was seen under parameters that should have yielded two steady CJ points. The detonation evolved to have a single CJ point instead of the two predicted in figure \ref{fig:Steady_state_2_CJ_points}.
The piston initiation transient was removed by initiating simulations with the predicted steady integral curves to investigate the discrepancy. However, none of the steady profiles prove to be stable, ending up with a CJ point B and subsonic point A (with respect to the detonation front, {i.e.} in the strong solution). The detonation front propagated slower than expected while under the influence of point A, shown to be a result of numerical diffusion.

Interestingly, the internal shocks, when created, always formed at sonic point B. This indicates sonic points A and B may be unstable when they coexist. The creation of internal shocks at sonic points will be investigated using the method of characteristics in section \ref{sec:Inspection of sonic points by the method of characteristics} and using linear stability analysis in section \ref{sec:Lin_stab_analysis}.

\subsection{\label{sec:Inspection of sonic points by the method of characteristics}Inspection of sonic points by the method of characteristics}
\begin{figure*}[]
	\centering %
	\begin{subfigure}[t]{0.23\textwidth}
		\includegraphics[width=\textwidth]{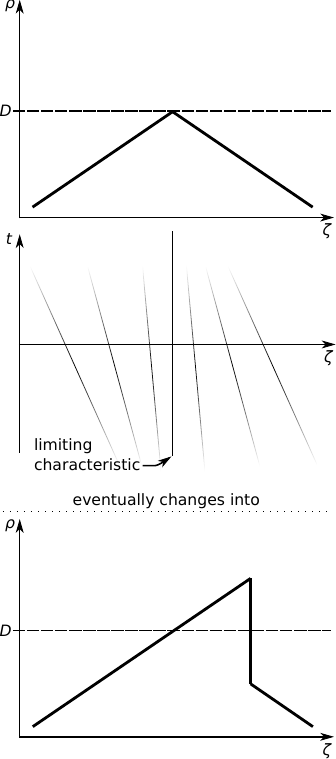}
		\caption{WW sonic point results in an upstream shock}
		\label{fig:non-linear_analysis:--}
	\end{subfigure} \ \ 
	\begin{subfigure}[t]{0.23\textwidth}
		\includegraphics[width=\textwidth]{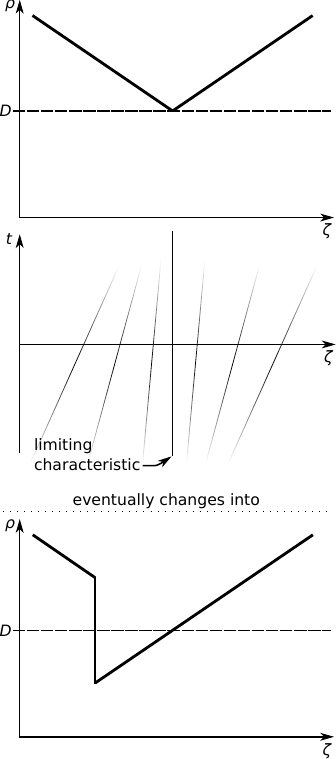}
		\caption{SS sonic point results in a downstream shock}
		\label{fig:non-linear_analysis:++}
	\end{subfigure} \ \ 
	\begin{subfigure}[t]{0.23\textwidth}
		\includegraphics[width=\textwidth]{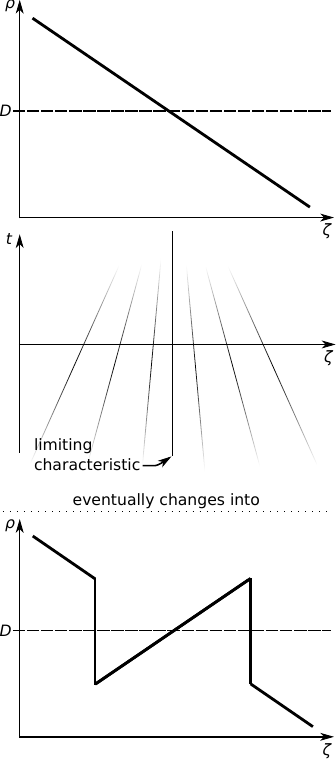}
		\caption{SW sonic point results in shocks on either side}
		\label{fig:non-linear_analysis:+-}
	\end{subfigure} \ \ 
	\begin{subfigure}[t]{0.23\textwidth}
		\includegraphics[width=\textwidth]{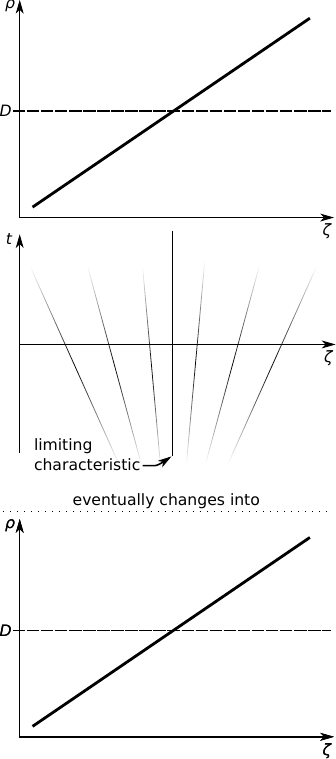}
		\caption{WS sonic point remains stable}
		\label{fig:non-linear_analysis:-+}
	\end{subfigure} \ \ 
	\caption{Illustration of shock creation in vicinity of a sonic point: initial signal speed $\rho$ (top), initial instantaneous characteristics (middle), and resulting signal speed $\rho$ (bottom)}
	\label{fig:non-linear_analysis}
\end{figure*}

In the following analysis, sonic points will be labelled in the same manner as the steady-state solutions: sonic points lying between two weak solutions ({e.g.} point B seen in the WWS structure of figure \ref{fig:piston_wave_profiles_piston:t=25.6311}) will be called a sonic point of type WW. Likewise, a sonic point between two strong solutions ($\rho > D$) will be named SS, WS will be used for sonic points with a strong upstream solution and weak downstream solution ($\rho < D$), and SW will be used when there is a weak upstream and strong downstream solution. The top frames of figure \ref{fig:non-linear_analysis} serve as illustrations.

Internal shocks were created in all simulations, with the exception of the WSS structure and the over-driven viscous SSS structure. Point B was always the source for these shocks except when of type WS. SS-type sonic points quickly became subsonic with respect to the detonation front, but WS sonic points remained stable until shocked. It appears WS sonic points are stable, unless shocked, while the others are unstable.

The method of characteristics was used near the two sonic points to assess their stability. Figure \ref{fig:non-linear_analysis} shows the flow in vicinity of the four types of sonic points. The top frames show $\rho$, which can be interpreted as the signal speed in Fickett's analogy, for the four types of sonic point. The bottom frames show the resulting profiles of $\rho$ expected from the simulations and following analysis. The middle frames show the instantaneous characteristics (information paths) corresponding to the top frames. Perturbations travel along these characteristic lines which each have a slope $\mathrm{d}t/\mathrm{d}\zeta \propto 1/(\rho-D$). The limiting characteristic travels at the same speed as the detonation front along the sonic point. For the strong solution, perturbations from the rear (left) are able to reach the limiting characteristic (and detonation front) as they travel faster than the front. For the weak solution, perturbations cannot keep up with the front and are therefore carried towards the left, away from the limiting characteristic.

For a sonic point of type WW (figure \ref{fig:non-linear_analysis:--}) the characteristic diagram shows that any perturbation downstream (left) of the sonic point will easily be dissipated to the rear as the information speeds away from the front in that direction. Upstream perturbations (to the right), will accumulate and eventually disturb the sonic point. 
If the perturbations cross the sonic line, a shock wave must be formed to link the weak upstream solution to the new strong solution immediately to the right of the sonic point. Flow immediately to the right of the sonic point now satisfies the strong solution while there is no change to the weak solution downstream. The WW sonic point transitions to a WS sonic point, seen in the bottom frame of figure \ref{fig:non-linear_analysis:--}, with an upstream shock that connects the strong solution to the weak solution. This behaviour is mirrored for SS sonic points, in opposite directions.

The sharp kink at the WW point may become temporarily curved and supersonic, seen at point B in figure \ref{fig:piston_wave_profiles_piston:t=25.6311}, allowing perturbations from the upstream side to slip downstream. This is mirrored in SS points which become temporarily subsonic, seen at point A of figures \ref{fig:wave_profiles_--+:end} and \ref{fig:wave_profiles_+-+:end}. These curved super/subsonic profiles are not solutions of the inviscid steady equations with two CJ points, but are caused by numerical (or physical) diffusion. The solutions with internal shocks shown in the bottom frames of figure \ref{fig:non-linear_analysis} are valid forms of the inviscid steady equations.

Sonic points of type SW, shown in figure \ref{fig:non-linear_analysis:+-}, are also unstable. Perturbations accumulate on either side, like they do on the right and left sides of WW and SS sonic points. Shocks are eventually created on either side of the sonic point which finally takes the WS form, as seen in figure \ref{fig:SWS}.

Finally, the characteristics around a type WS sonic point (figure \ref{fig:non-linear_analysis:-+}) show that perturbations should easily be dissipated away on either side. This explains the stability of this type of sonic point seen in the simulations.

This analysis suggests that only WS sonic points are stable due to their diverging characteristic paths. The other types of steady sonic point should transform to create shocks and stable WS sonic points. This leads to a further consequence: in order for a solution with two coexisting sonic points to remain steady and stable, the sonic points must be separated by a shock wave. The location of this shock between the two sonic points is arbitrary in the framework of this model. Although this internal shock is protected from disturbances behind the second sonic point, it is affected by disturbances originating between the sonic points. In simulations, internal shocks between the sonic points always reached the detonation front. This instability is ascribed to the effect of numerical diffusion and a feature of the model, whereby the internal shock may exist anywhere between two sonic points.

It is not clear how this translates to more complex models (such as the Euler equations) and more realistic descriptions of kinetic rates, where shocks may be anchored by the state dependence of reactions. Steady internal shocks between CJ points have been seen by Veyssi\`{e}re and Khasainov \cite{veyssiere1991model,veyssiere1995structure}. While investigating these double-front detonations, they found \cite{khasainov1996initiation} that ``the distance between the two subsonic zones evolves very slowly'' at the end of their unsteady simulations. The internal shock in the Euler equations may also be unstable, but further study is required.

\subsection{\label{sec:Lin_stab_analysis}Linear stability analysis of sonic points}

The stability of the sonic points can also be studied using linear stability analysis. A new independent variable $s$ is defined such that
$\text{d} \zeta / \text{d} s = \rho - D$
in order to remove the singularity from equation \ref{eq:SS_ODE_density}, yielding a system of equations
\begin{align}
\label{eq:ODE_singularity_free}
\frac{\text{d} \zeta}{\text{d} s} = \rho - D 
\text{, } &&
\frac{\text{d} \rho}{\text{d} s} = \frac{\sigma}{D} \text{,}
&& \text{ and } &&
\frac{\text{d} \lambda_i}{\text{d} s} = -\frac{r_i}{D} (\rho-D).
\end{align}
Setting these equations to zero reveals fixed points where the generalized CJ criterion of equation \ref{eq:Generalized_CJ_condition} is met at sonic points A and B, {i.e.} $\rho = D$ and $\sigma = 0$. Linearising the singularity-free system (equation \ref{eq:ODE_singularity_free}) around sonic point A, whose reaction progress variables are in equation \ref{eq:CJ_A}, gives the Jacobian of $[\zeta, \rho, \lambda_1,\lambda_2,\lambda_3]$
\begin{align*}
J_{\text{A}} &= 
\left[\begin{array}{c c c c c}
0 & 1 & 0 & 0 & 0
\\ 0 & 0 & \begin{matrix} - \frac{k_1 Q_1 \nu_1}{2 D} \left(-\frac{k_3 Q_3}{k_1 Q_1}\right)^{\frac{\nu_1-1}{\nu_1}} \end{matrix} & 0 & 0
\\ 0 & - \frac{k_1}{D} \left(-\frac{k_3 Q_3}{k_1 Q_1}\right)  & 0 & 0 & 0
\\ 0 & 0  & 0 & 0 & 0
\\ 0 &  - \frac{k_3}{D} & 0  &  0 & 0
\end{array}\right]
\end{align*}
with eigenvalues $e_{\text{A}}=\pm \sqrt{\frac{k_1^2 Q_1 \nu_1}{2 D^2}  \left(-\frac{k_3 Q_3}{k_1 Q_1}\right)^{\frac{2\nu_1-1}{\nu_1}}}$ and $e_{\text{A}}=0$. All parameters being positive, with the exception of $Q_3$, the two non-zero eigenvalues are real and of opposite signs. A saddle node \cite{strogatz2014nonlinear} therefore exists at point A with eigenvectors
\begin{align*}
E_{\text{A}} = \left[\frac{1}{e_{\text{A}}}, 1, -\frac{k_1}{D}\left(\frac{-k_3 Q_3}{k_1 Q_1}\right) \frac{1}{e_{\text{A}}}, 0, \frac{-k_3}{D} \frac{1}{e_{\text{A}}} \right] ^T .
\end{align*}
This analysis at point B yields the same eigenvalues, with parameters of the first reaction replaced by those of the second, and same for the eigenvectors.

\begin{figure}[]
	\centering %
	\includegraphics[scale=1]{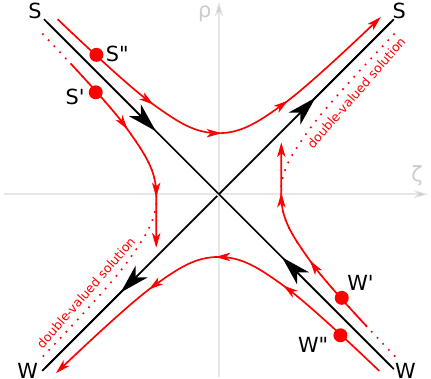}
	\caption{Illustration of the linear stability analysis around the sonic points in the $\zeta$-$\rho$ plane, revealing the existence of a saddle node; typical points S and W (strong and weak) lying on the detonation profile are shown with their trajectories in red; single prime labels points between the eigenvectors, double prime for points above and below}
	\label{fig:stability}
\end{figure}

The saddle node is illustrated in figure \ref{fig:stability} in $\rho$-$\zeta$ space. The unstable eigenvector with a positive slope and the stable eigenvector with a negative slope are drawn in black. The typical trajectories of signals are plotted in red. 

Consider a sonic point of type SS which is perturbed on the left side to have a value of $\rho$ slightly above its ideal value, it can be written as S$''$S in accordance with the figure. The signal from the point S$''$ will first travel towards the saddle node then be repulsed and move away from the node in the S direction on the right side. This gives rise to a curved subsonic point like seen at A in figures \ref{fig:piston_wave_profiles_piston:end}, \ref{fig:wave_profiles_+-+:end}, and \ref{fig:wave_profiles_--+:end} whose perturbation may be caused by rounding through diffusion. The same process works in the reverse direction for a type WW$''$ point.

A signal S$'$ of the strong branch starting below the eigenvector on the left side will travel towards the node, but following its trajectory will eventually force a double-valued $\rho$ which is inadmissible along the $\zeta$ axis so a shock will form. The story is the same for a signal originating from the W$'$ position.

The linear stability analysis mirrors the conclusions made using the method of characteristics.
Disturbances from the bottom left or top right are repulsed from the node and do not cause shocks, so WS sonic points remain stable. Sonic points of type S$''$S and WW$''$ will become non-sonic, and points with S$'$ or W$'$ will create shocks.

\section{\label{sec:Conclusion}Conclusion}

Detonations with two sequential heat releases and a concurrent loss were studied using Fickett's detonation analogue. The study aimed to clarify the steady reaction zone structure and velocity when the two sonic planes propagated with different speeds (the general case). It was found that a steady detonation must travel at the maximum velocity dictated by the net heat release between the front and each zero-thermicity point. Singularities were found in the flow structure at lesser velocities, and the lack of sonic point at larger velocities opened the detonation front to perturbations. This supports other studies \cite{bdzil2006higher,veyssiere1991model,khasainov1996initiation,veyssiere1995structure} which found transitions from the lower detonation velocities to higher ones, but not vice-versa.

Piston-initiated unsteady numerical simulations showed that the lower detonation velocities were quasi-steady. The formation of a forward-travelling shock and its subsequent arrival at the front caused the detonation to accelerate to its higher velocity. This phenomenon is believed to be the cause of the detonation front's sudden acceleration previously mentioned by Bdzil {et al.} \cite{bdzil2006higher}.

The second goal of this study was to determine the stability of the internal shock sometimes present in double-structured detonations. This was done by studying cases where both sonic planes coexisted. The presence of one sonic point in the flow allows the existence of an internal shock behind. Internal shocks were found to be unstable due to diffusive effects and to the independence of reaction rates from the flow field, which allows internal shocks to exist anywhere. It is not clear if or how reaction dependence on hydrodynamics could anchor internal shocks, further work is needed to establish in what capacity internal shocks can be stable, as suggested by Veyssi\`{e}re and Khasainov \cite{veyssiere1991model,veyssiere1995structure}.

Finally, the origin of the internal shock was sought. Perturbations were found to accumulate at any sonic point with characteristics that do not diverge on both sides, leading to the creation of internal shocks in many cases. This is supported by a linear stability analysis which clarifies why perturbations form shocks or rounded points. As a consequence, two sonic points may only co-exist in a stable fashion (WS) if they are separated by a shock. This clarifies the roots of double front detonations seen in previous work \cite{zhang2006detonation,veyssiere1995structure, zhang2009detonation,afanasieva1983multifront,veyssiere1984double}, and is also the mechanism by which detonations transition from slower quasi-steady velocities to faster ones.

\section*{Acknowledgements}
	The authors would like to acknowledge funding from the Defence Research and Development Canada (DRDC), the Ontario Graduate Scholarship (OGS) fund, as well as the Natural Science and Engineering Research Council (NSERC) of Canada in the form of a Canadian Graduate Scholarship (CGS).

\bibliographystyle{tfq}
\bibliography{references}

\end{document}